\documentclass[aps,prl,10pt,twocolumn,showpacs,preprintnumbers,amsmath,amssymb,superscriptaddress]{revtex4-1}

\usepackage[caption=false]{subfig}
\usepackage{lineno}
\usepackage{graphicx}  
\usepackage{dcolumn}   
\usepackage{bm}        
\usepackage{amssymb}   
\usepackage{amsmath}
\usepackage{blkarray, multirow, graphicx, diagbox, color, xcolor, colortbl}
\usepackage{bbm, bbold}
\usepackage{ifthen}
\usepackage[colorlinks, linkcolor = blue, citecolor = blue, filecolor = black, urlcolor = blue]{hyperref}
\usepackage{xkeyval}
\usepackage{moreverb}
\usepackage{rotating}
\usepackage{slashbox}
\usepackage{xspace}
\usepackage{nicefrac}
\usepackage[]{units}
\usepackage[]{natbib}
\usepackage{physics}
\usepackage{braket}
\usepackage{hyphenat}
\usepackage{amsbsy}

\usepackage{ulem}

\begin{document}

\title{Ultrafast Optical Control of Magnetic Order and Fermi Surface Topology at a Quantum Critical Point}

\author{Benedikt Fauseweh}
\email{fauseweh@lanl.gov}
\affiliation{Theoretical Division, Los Alamos National Laboratory, Los Alamos, New Mexico 87545, USA}

\author{Jian-Xin Zhu}
\email{jxzhu@lanl.gov}
\affiliation{Theoretical Division, Los Alamos National Laboratory, Los Alamos, New Mexico 87545, USA}
\affiliation{Center for Integrated Nanotechnologies, Los Alamos National Laboratory, Los Alamos, New Mexico 87545, USA}

\date{\today}

\begin{abstract}
Designing material properties on demand has important implications to potential future technological applications. While theoretically it is always possible to tune various intrinsic energy scales, there are fundamental limitations to this approach in experiment. In recent years, ultrafast spectroscopy has evolved as a promising tool to use light to dynamically induce non-trivial electronic states of matter. Here we theoretically investigate  light pulse driven dynamics in a Kondo system close to quantum criticality. We show, that light can dehybridize the local Kondo screening and induce magnetic order out of a previously paramagnetic state.  We demonstrate that, depending on the laser pulse field parameters, it is possible to deconfine the Kondo singlet and thereby induce second order phase transition to a dynamically ordered state, as well as a dynamic Lifshitz transition that changes the Fermi surface topology from hole- to electron-like. 
\end{abstract}


\maketitle

\section{Introduction}

\begin{figure*}
\includegraphics[width=1.0\textwidth]{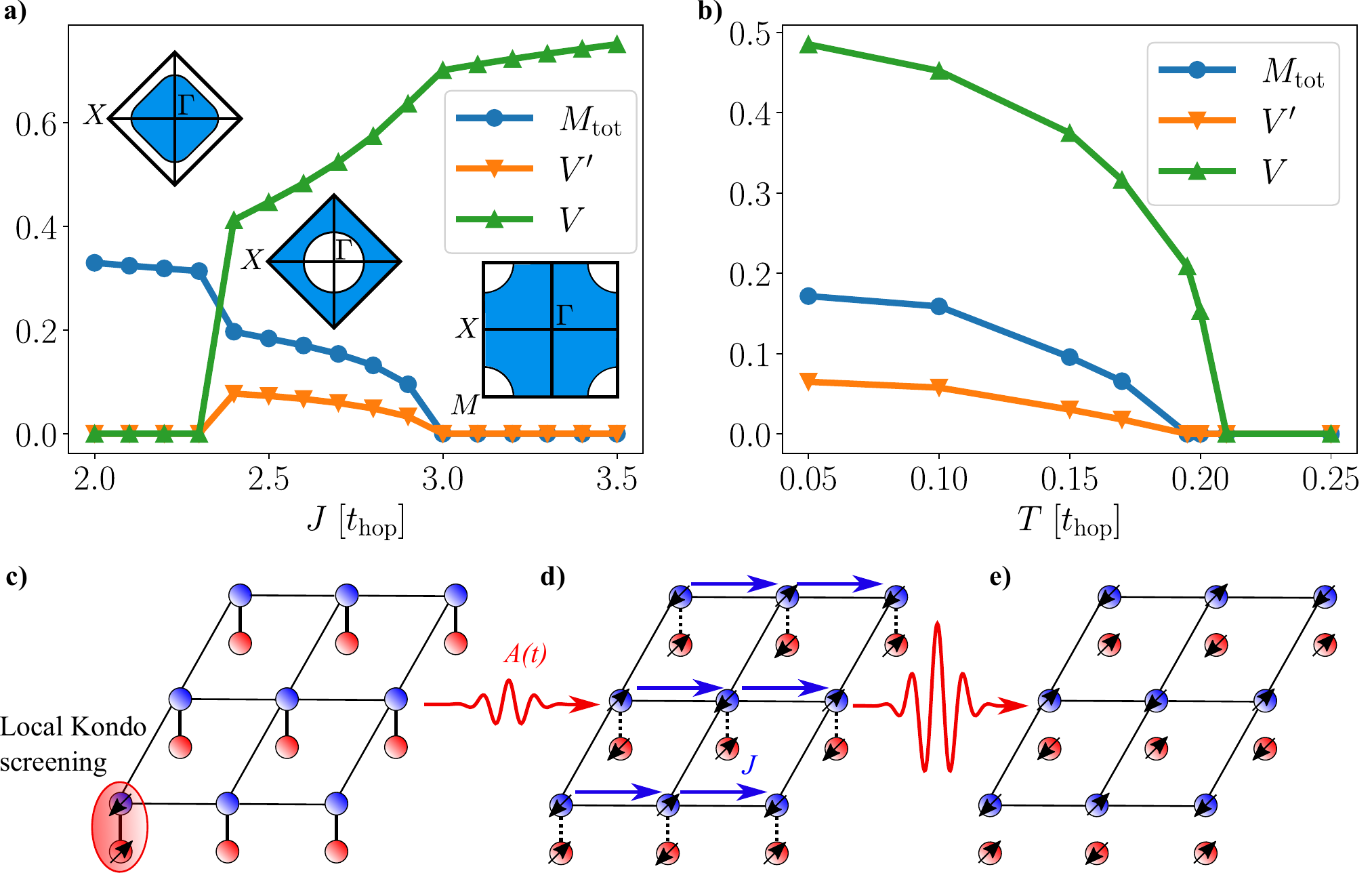}
\caption{  \textbf{a)} Equilibrium expectation values for total magnetization $M_\mathrm{tot}$, total hybridization $V$ and staggered hybridization $V'$, as function of Kondo interaction $J$ at temperature $T=0.01 t_\mathrm{hop}$. We destinguish three different phases: 1) The PM phase for strong Kondo coupling has a large BZ with hole-like  pockets at the M points. 2) Intermediate regime with finite magnetization.   The FS remains topologically invariant but is folded into the smaller BZ. 3) Small Kondo coupling regime leads to a decoupling between the charge carriers and the local magnetic moments. The Fermi surface topology changes from hole- to electron-like.  \textbf{b)} Same as in \textbf{a)} for $J=2.7 t_\mathrm{hop}$ as function of temperature $T$.  \textbf{c)} Ground state representation of the Kondo lattice model in the PM Kondo screened phase.  \textbf{d)} Light-induced magnetization accompanied with a weakening of the Kondo screening and induced current. \textbf{e)} Further increasing the pulse amplitude cancels the Kondo screening.  }
\label{Fig1}
\end{figure*}

\begin{figure*}
\includegraphics[width=1.0\textwidth]{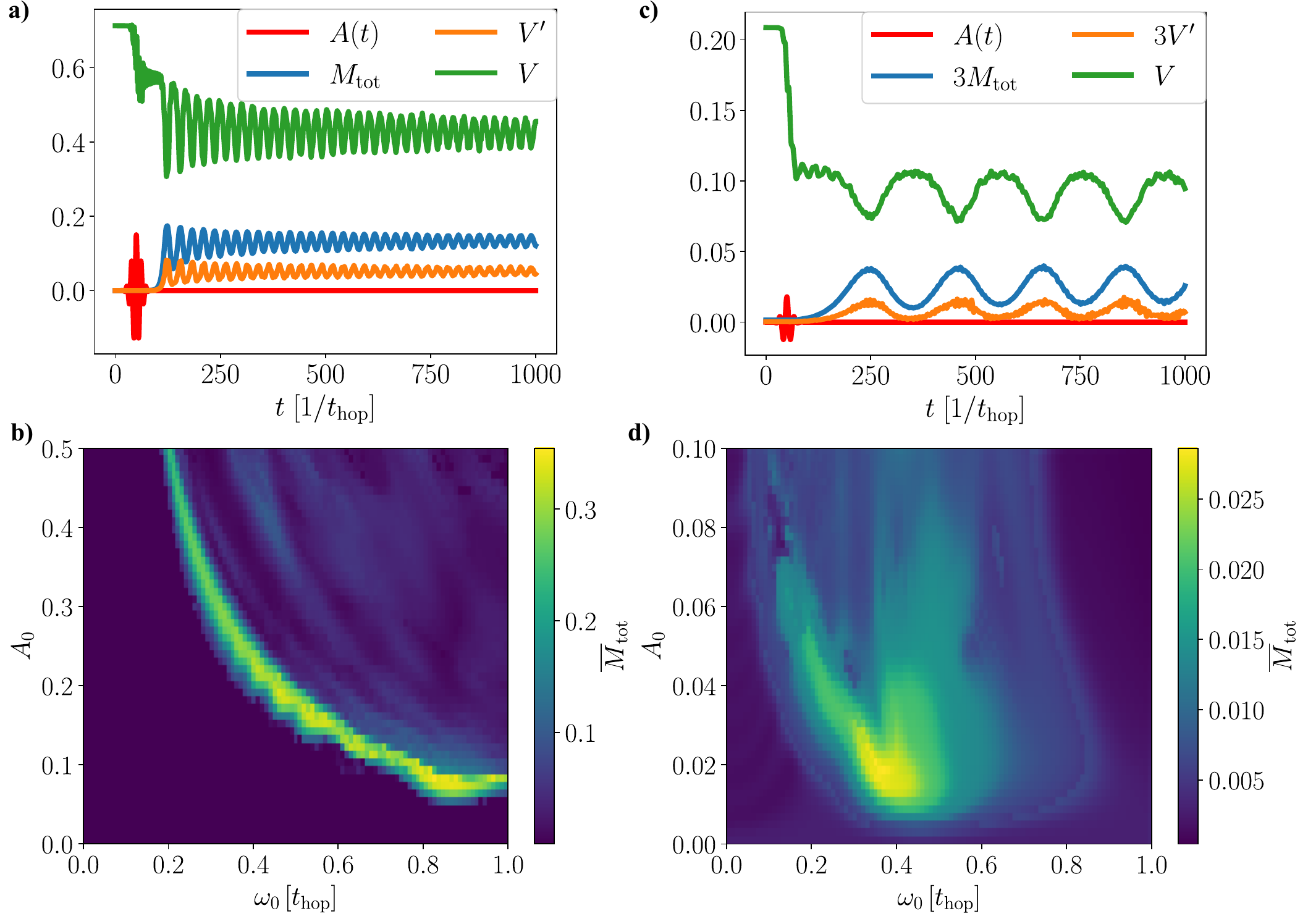}
\caption{\textbf{a)} Time dependence of magnetization and hybridization after a short light pulse in the PM phase at $J=3.1 t_\mathrm{hop}$ and $T=0.01 t_\mathrm{hop}$. \textbf{b)} Dynamical Phase Diagram of the time averaged magnetization $\overline{M}_\mathrm{tot}$ after light-pulse as function of pulse amplitude $A_0$ and frequency $\omega_0$ at $J=3.1 t_\mathrm{hop}$ and $T=0.01 t_\mathrm{hop}$. \textbf{c)}   Same as in \textbf{a)} but in the mixed phase with $J=2.7 t_\mathrm{hop}$ and $T=0.195 t_\mathrm{hop}$. \textbf{d)} same as in \textbf{b)} but in the mixed phase with  $J=2.7 t_\mathrm{hop}$ and $T=0.195 t_\mathrm{hop}$. 
  }
\label{Fig2}
\end{figure*}
 
 \begin{figure*}
\includegraphics[width=1.0\textwidth]{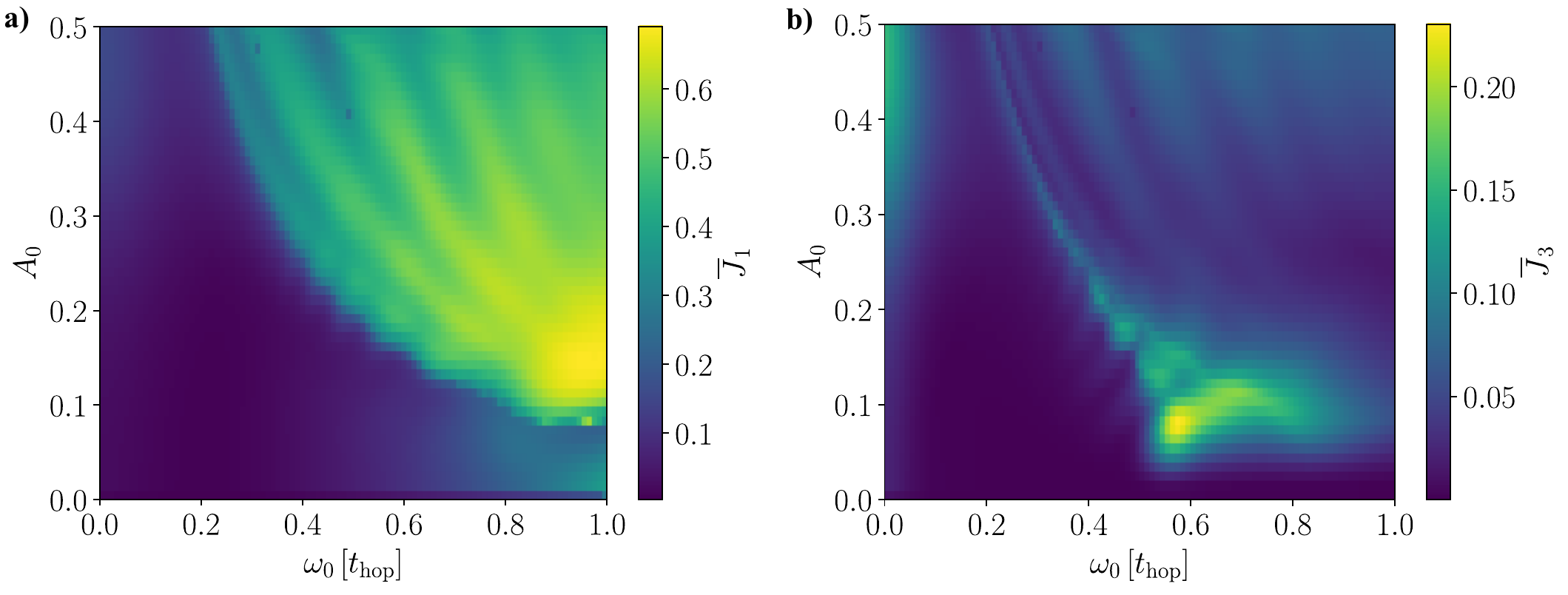}
\caption{Intensity of the fundamental  harmonic \textbf{a)} and third harmonic \textbf{b)} after a short light pulse in the PM phase at $J=3.1 t_\mathrm{hop}$ and $T=0.01 t_\mathrm{hop}$.  }
\label{Fig3}
\end{figure*}
 
 \begin{figure*}
\includegraphics[width=1.0\textwidth]{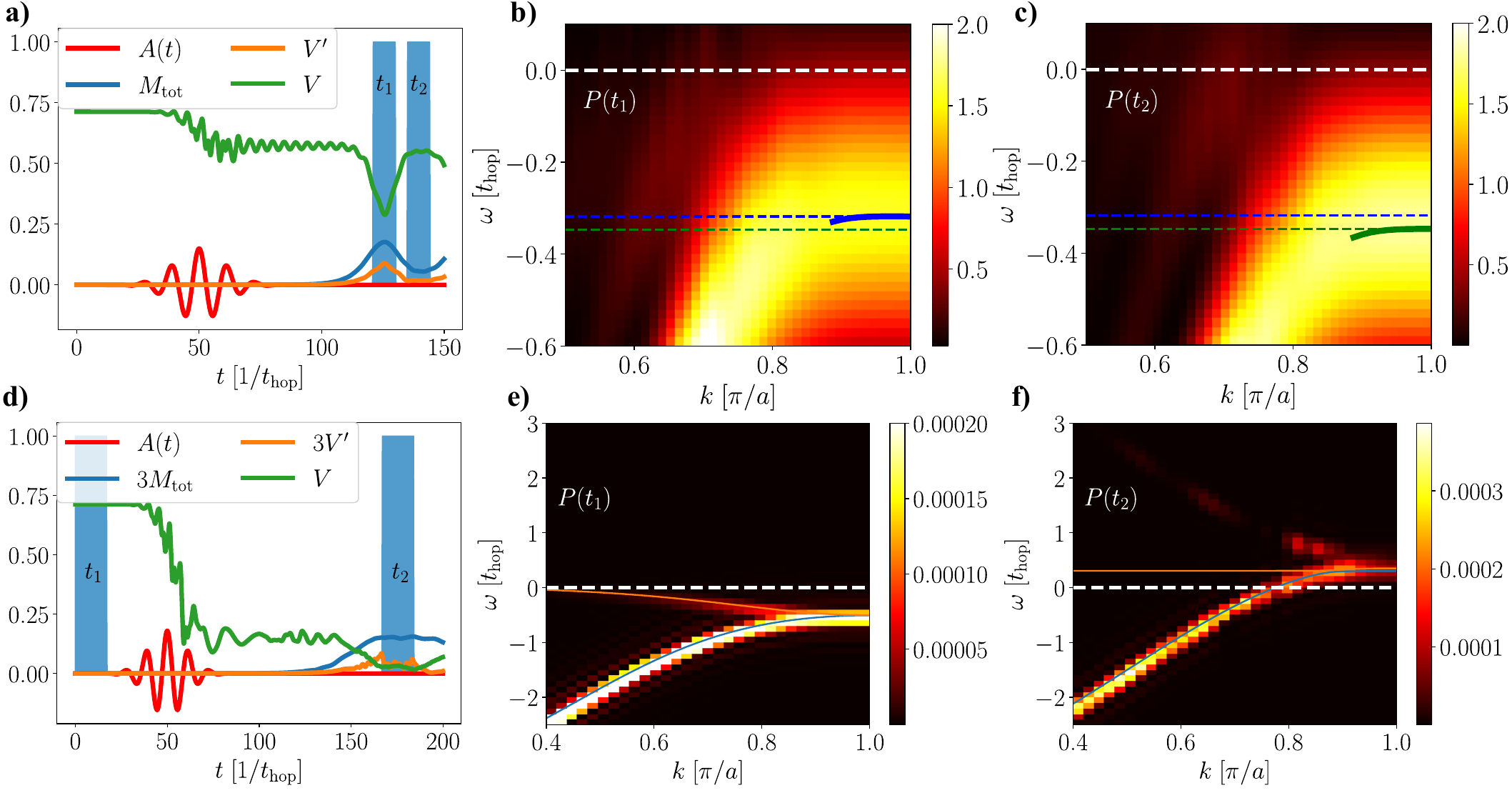}
\caption{\textbf{a)} Dynamics after pump pulse inducing order parameter oscillations or $A_0=0.15$ and $\omega_0=0.55 / t_\mathrm{hop}$. Shaded regions show the time domain for the photoemission spectroscopy probe pulses. \textbf{b) and c)} Photoemission spectra measured at $t_1$ and $t_2$ respectively, as shown in \textbf{a)}. Solid lines show the maximum of the signal as function of $k$ while dotted lines are for comparison. \textbf{d)} Dynamics for stronger pump pulse that significantly reduces the hybridization. Pulse parameters are $A_0=0.18$ and $\omega_0=0.55 / t_\mathrm{hop}$. \textbf{e)} Photoemission spectra measured at $t_1$ before the pump pulse as shown in \textbf{d)}. Solid lines are the time averaged lowest eigenvalues of the mean-field Hamiltonian. \textbf{f)} Photoemission spectra measured at hybridization minimum at $t_2$ after the pump pulse in \textbf{d)}.  Solid lines as in \textbf{e}). All momenta measured in $k = \left| (k_x, k_y) \right|$ for a path along the diagonal in the reduced BZ.  }
\label{Fig4}
\end{figure*}

Quantum materials exhibit non-universal properties that are the result of strong interactions between their constituents, defying  a theoretical description in terms of free electrons. Their physical phenomena, such as superconductivity, long range entanglement and topologically protected currents, are exotic and absent in simple metals or insulator \cite{Keimer2017}. This makes them promising candidates for applications in future quantum devices.

Classifying and predicting such unusual states of matter and their excitation spectra in terms of broken symmetries and topological properties is one of the central aims in modern condensed matter physics. These efforts lead to the notion of quantum critical points \cite{Sachdev,Coleman2015,Si1161} (QCPs), which are zero temperature phase transitions separating different ground states driven by quantum fluctuations. In many of the currently investigated quantum materials, the competition between itinerant dynamics and the tendency to localization, induced by the Coulomb interaction between the electrons,  is the driving force for strong correlations and the formation of QCPs in quantum materials. 

In heavy fermion systems, such as Ce-, and Yb-based compounds \cite{RevModPhys.73.797,RevModPhys.79.1015,Gegenwart2008}, localized $f$ electrons  act as magnetic moments, that are interacting with conduction band electrons via the Kondo interaction, leading to the formation of local Kondo singlets. The Ruderman-Kittel-Kasuya-Yosida (RKKY) interaction, which is generated from the Kondo interaction but mediated via the conduction band electrons, is competing with this local Kondo singlet formation and favors magnetic ordering. In the Kondo limit, i.e. large Kondo interaction $J$, the local singlet formation leads to a Kondo screening effect of the conduction band electrons. No magnetic order exists and the system is paramagnetic. Due to the strong hybridization between $f$- and conduction band electrons, the effective mass of the low-energy band electrons gets strongly renormalized and the Fermi surface (FS) volume expands, providing a signature of the Kondo-effect induced heavy Fermi-liquid phase, see Fig.\ 1 a) (large FS). In contrast if the RKKY interaction dominates, the system exhibits magnetic order and the Kondo heavy quasiparticles are destroyed. The transition between these two phases is non-trivial, as in the conventional picture the conduction and $f$ electrons would always hybridize, leading to heavy fermions on both sides of the QCP \cite{PhysRevB.14.1165,PhysRevB.48.7183,doi:10.1143/JPSJ.64.960}. However, in many experiments a different behavior is observed \cite{Coleman_2001,RevModPhys.73.797,L_hneysen_1996,Schroder2000,Guttler2019}, in which a reconstruction of the FS takes place abruptly and the Kondo screening becomes irrelevant. In this case the FS is only determined by the conduction band electrons (small FS). This scenario is referred to as local quantum criticality \cite{Si2001,Gegenwart2008,RevModPhys.92.011002}.

While traditional methods to unravel these interesting phenomena in experiment include chemical doping, application of static pressure and magnetic fields \cite{Wirth2016}, in recent years it has become possible to use laser pulse femtosecond drives to dynamically induce novel states in quantum materials \cite{Basov2017}. With this approach it was possible to observe transient superconductivity in various materials \cite{Fausti189,Mankowsky2014,Mitrano2016,PhysRevB.101.180507,PhysRevX.10.031028}, ultrafast switching of Weyl semimetals \cite{Sie2019}, magnetization dynamics in van der Waals magnets \cite{Seyler2018,Sun2019,padmanabhan2020coherent} as well as Higgs oscillations in unconventional superconductors \cite{matsunagaPRL13,Matsunagascience14,Schwarz2020,Chu2020}. 
The search for new light induced quantum phases, is a central aim in modern solid state physics \cite{Kennes2017,Yang2018, Giorgianni2019}.
Also heavy fermion systems have been investigated in ultrafast experiments \cite{Wetli2018,PhysRevLett.104.227002,PhysRevB.97.165108,PhysRevLett.122.096401}, calling for a theoretical description of the interplay between quantum criticality and non-equilibrium dynamics in these system.

In this paper we theoretically investigate, how short laser pulses can influence the local Kondo screening and magnetic order in the Kondo square lattice geometry. We use time-dependent mean-field theory supported by time-dependent Variational Monte Carlo calculations to demonstrate, that a dynamical quantum phase transition from a pure paramagnetic phase to an oscillating antiferromagnetic state can be induced, see Fig. 1 d). We analyze the effect of laser pulse amplitude as well as frequency to map out a dynamical phase diagram (DPD) and show, that a strong laser pulse can dynamically suppress the hybridization between local magnetic moments and conduction band electrons  see Fig. 1 e), leading to sudden change of the FS from hole- to electron-like, i.e. a transient Lifshitz transition in the topology of the FS. We predict, that high harmonic generation (HHG)~\cite{PhysRevB.102.165128}, as well as time-resolved angle-resolved photoemission spectroscopy (tr-ARPES) are highly sensitive tools to probe these phenomena.
 
\section{Results}

To analyze the interplay between Kondo screening and non-equilibrium dynamics we use the square lattice Kondo model as a prototypical example \cite{PhysRevB.87.205144,PhysRevLett.99.136401}. Under the assumption of a low frequency pulse, we can safely neglect the local $f$-interband transitions and purely work with a local magnetic moment degree of freedom.  We focus on the uncompensated, metallic regime at $n_c = 0.9$, where $n_c$ is the conduction band filling, and perform a time-dependent mean-field analysis. In equilibrium, three different phases can be distinguished depending on Kondo interaction strength at zero temperature, see Fig. 1 a). In the strong Kondo limit the system is paramagnetic, with finite hybridization between the conduction band electrons and the local magnetic moment. As the Kondo interaction is reduced the RKKY interaction starts to compete with the local Kondo singlet formation and the system crosses a second order phase transition at $J \approx 3 t_\mathrm{hop}$ into a mixed phase, where hybridization as well as antiferromagnetic order coexist. The total antiferromagnetic magnetization is denoted as $M_\mathrm{tot} = m_c - m_f$, where $m_c$ and $m_f$ are the conduction- and local moment magnetizations respectively, while $V$ denotes the total hybridization. Note that only in the intermediate phase, the staggered hybridization, i.e. $V' = V_A - V_B$, where $V_{A/B}$ is the hybridization on the $A/B$ sublattice, is finite. Upon further reducing the Kondo interaction a first order transition at $J \approx 2.4 t_\mathrm{hop}$ completely destroys the hybridization between conduction and local moments and the FS changes its topology from hole- to electron-like, marking a Lifshitz transition, i.e. a sudden reconstruction of the Fermi surface giving rise to anomalies in the electronic characteristics of metals. 
Within the intermediate regime, i.e. $J=2.7 t_\mathrm{hop}$, we investigate the finite temperature phase in Fig. 1 b). Thermal fluctuations affect the magnetic order first, with a Neel temperature of about $T_N \approx 1.9 t_\mathrm{hop}$, while the Kondo temperature is  $T_K \approx 2.1 t_\mathrm{hop}$. Note that the finite Neel temperature is due to the mean-field approximation in our calculation. In an exact treatment it would be at zero temperature and we assume that some additional mechanism, such as spin interaction anisotropy, leads to a finite Neel temperature.
Starting within the PM Kondo screened phase, we investigate the effect of a finite laser pulse onto magnetization and hybridization. We parameterize the light pulse using a Gaussian envelope $A(t) = A_0 \exp(-(t-t_c)^2/2 t_d^2) \cos(\omega_0 (t-t_c))$, where $t_c = 50 / t_\mathrm{hop}$ is the pulse center, $t_d = 10  / t_\mathrm{hop}$ is the pulse width, $A_0$ is the overall pulse amplitude and $\omega_0$ is the pulse frequency. Figure 2 a) shows an example of the time evolution of the system as well as the time dependence of the light vector potential. During the laser pulse the Kondo screening is reduced, signaled by the reduction of the hybridization between local moments and conduction band electrons. At this point in time the RKKY interaction becomes the dominant energy scale and induces the formation of long range magnetic order directly after the pulse. The structure of this magnetic order is identical to the intermediate equilibrium regime but the order parameter oscillates around its asymptotic value $\bar{M}_\mathrm{tot}$.  These order parameter oscillations are similar to the recently observed Higgs oscillations in superconducting condensates. Similar to the Higgs oscillations, the oscillation frequency is proportional to the asymptotic order parameter value. At the same time we observe hybridization oscillations with the same frequency but with a $\pi$-phase shift. A finite staggered component $V'$  dynamically evolves after the pulse, that has no phase shift with respect to  the antiferromagnetic order oscillations. We observe only a very slow decay of the order parameter oscillations, indicating the formation of a stable collective magnetic excitation. To quantify the dependence of the induced magnetization on the pulse parameters, we investigate the DPD of the asymptotic magnetization in Fig. 2 b) as function of pulse amplitude and frequency at fixed width and center. In order to induce magnetization, pulse frequency $\omega_0$ as well as amplitude $A_0$ have to be sufficiently large. Once a certain threshold value is reached we observe a dispersive region in which the induced magnetization is maximized. Further enhancing the pulse amplitude does not result in an increased magnetization but instead heats the system, melting the induced magnetization again.  Within this region, we can still induce magnetic order parameter oscillations, however their asymptotic value is close to zero. 
The possibility to dynamically induce magnetic order out of an unordered state is not limited to the zero temperature case. In Fig. 2 c) we investigate the system dynamics within the intermediate regime, i.e. $J= 2.7 t_\mathrm{hop}$, but above the Neel temperature at $T= 0.195  t_\mathrm{hop}$, still within the Kondo screened phase. After the initial hybridization drop, we observe the dynamical formation of magnetic order. The overall amplitude of the induced magnetization is reduced when compared to the zero temperature case, resulting in slower order parameter oscillations. The corresponding finite temperature DPD of the asymptotic magnetization is shown in Fig. 3 d). It shows a much broader region when compared to the zero temperature case.

To identify the dynamical phase transition in experiment we investigate the HHG spectrum induced by the pump pulse. Specifically we concentrate on the intensity profiles of the fundamental and third harmonic as a function of pulse amplitude and frequency in Fig. 3 for the zero temperature case. The fundamental harmonic shows a strong response upon entering the Kondo breakdown regime, see for comparison Fig 2. b). A fine structure is visible within the spectra. By changing the pulse width $\tau_d$ we can vary this fine structure, therefore we attribute it to the finite pulse width in the simulation. The third harmonic is also sensitive to the magnetic phase transition. Additionally it features a strong response for $A_0 \approx 0.07 t_\mathrm{hop}$ and $\omega_0 \approx 0.58 t_\mathrm{hop}$,  just before the actual breakdown of the Kondo screening at stronger pump pulses. 
Using only HHG spectroscopy we can pinpoint the transition region but it is insufficient to directly investigate the microscopic non-equilibrium physics. We thus use the method introduced by Freericks, Krishnamurthy and Pruschke \cite{Freericks2009} to compute tr-ARPES spectra. Specifically we compute the PES signal $P(t, \omega, k) = \int \tau_1 \int \tau_2 s_t(\tau_1) s_t(\tau_2) e^{i \omega (\tau_1 - \tau_2)} G^{<} (k, \tau_1, \tau_2)$, where $s_t(\tau)$ is the probe pulse shape function centered around $t$ and $G^{<}$ is the two-time lesser Green function. Fig 2 a) shows the probe pulse as shaded regions, parameterized as
\begin{align}
s_t(\tau) = \begin{cases} 1 &\mbox{if } t-t_w/2 < \tau < t+t_w/2 \\
e^{-(\tau-(t-t_w/2))^2/ 2 t_s^2} &\mbox{if }  \tau \leq t-t_w/2 \\
e^{-(\tau-(t+t_w/2))^2/ 2 t_s^2} &\mbox{if }  \tau \geq t+t_w/2  \end{cases}
\end{align}
where $t_w$ is the width of the probe pulse, and $t_s$ is the switching time. To resolve the light pulse induced dynamics directly we need a pulse that is short enough to not average over the dynamics, while being long enough to obtain a sufficient energy resolution. 
Figure 4 b) shows the ARPES signal centered around $t_1$ in Fig. 4 a), and mark in blue the maximum of the signal as function of momentum $k$. For comparison, Fig. 4 c) shows the same quantity centered around $t_2$ (marked in Fig. 4 a)), and highlight in green the maximum of the signal. The magnetic order oscillations directly manifest in oscillation of the spectral function in energy and amplitude. Also for $k < 0.75$ the oscillation affects the effective energy bands in terms of a small shift in  the chemical potential. The observation of both amplitude and energy oscillations has also been proposed to observe superconducting Higgs modes in tr-ARPES type experiments \cite{ PhysRevB.92.224517,PhysRevB.96.184518,PhysRevB.99.035117,PhysRevB.101.224510}.
To further elucidate the effect of stronger pump pulses we investigate a case, in which the Kondo screening is more strongly affected. Figure 4 d) shows the dynamics in which the hybridization is suppressed by up to $90 \%$ at $t_2$ during the time evolution. We compare this situation to the equilibrium probe before the pump pulse at a probe region around $t_1$. Figure 4 e) shows the ARPES signal at $t_1$ in equilibrium. Here only a single conduction band is present, folded into the smaller BZ for convenience. The FS is large and hole like within the small BZ. After optical excitation the situation changes drastically, as shown in Fig. 4 f). The induced magnetization splits the conduction band and strongly renormalizes the electronic structure. The FS is now significantly reduced (small FS) and we see excitations in the upper, previously empty, conduction bands. This change in the FS is similar to what is observed in equilibrium when the Kondo destruction induces a topological Lifshitz transition and a complete dehybridization of the conduction band electrons and local moments. While the equilibrium transition is triggered by the change of the Kondo interaction strength, here it is driven by fluctuations induced through the strong laser light.

\section{Conclusion}

The main result of our work is the prediction of a rare light-induced symmetry breaking in heavy fermion systems accompanied with the destruction of the Kondo coherence. We showed, that two different types of phase transitions, a second order transition with the appearance of magnetic order and a topological Fermi surface reconstruction can be dynamically driven using ultra-short laser pulses. We showed that the laser-induced regime is highly sensitive to the laser pulse parameters  and that HHG can be used to dynamically map out the phase transition region. We also predict, that tr-PES can observe magnetic order parameter oscillations, as well as the Lifshitz transition. These phase transition also have a significant effect on transport properties, such as the transient Hall coefficient, which should show sign change after the topological transition.
Our results indicate that the localized character of the Kondo PM phase is disturbed by the light field, while the formation of magnetic order via the RKKY interaction requires stronger light fields in order to be suppressed. Note that this selectivity is contrary to the effect of pure heating, as seen in Fig. 1 b), where temperature fluctuations induce a Neel temperature that is smaller than the Kondo temperature. 
The dynamically induced properties clearly show that the interplay between light, Kondo effect and magnetic order can have profound effects for heavy fermion materials.

\section{Methods}

\paragraph{Model}
The Kondo lattice model is defined by
\begin{align}
H = -t_\mathrm{hop}  \sum\limits_{\langle i , j \rangle, \sigma} \left( c_{i \sigma}^\dagger c_{i \sigma}^{\phantom\dagger} + \mathrm{H.c.} \right) + J \sum\limits_{i} \mathbf{S}_i^\mathrm{c} \cdot \mathbf{S}_i^\mathrm{f} .
\end{align}
The first term is the kinetic energy of the conduction electrons; the second term is the Kondo coupling of the conduction band to the local $SU(2)$ moments $\mathbf{S}_i^\mathrm{f}$ of the $f$-electrons. In the following we use $e=c=\hbar=a=1$, where $a$ is the lattice spacing. Unless otherwise noted, we use the hopping $t_\mathrm{hop}$ as our unit of energy and $1/t_\mathrm{hop}$ as our unit of time. \\
We include a time-dependent EM field by the well established Peierls substitution \cite{Claassen2017,Konstantinovaeaap7427,PhysRevLett.112.176404}
\begin{align}
t_\mathrm{hop} \rightarrow t_\mathrm{hop} e^{i \mathbf{A}(\tau) (\mathbf{r}_i - \mathbf{r}_j)  },
\end{align}
where $\mathbf{A}(t)$ is the time-dependent vector potential. 

\paragraph{Mean-field description}
We use a fermion description of the local magnetic moments $S_i^\alpha = 1/2 \sum_{\sigma, \sigma'} f_{i, \sigma}^\dagger \tau^\alpha_{\sigma, \sigma'} f_{i,\sigma'}$ leading to the purely fermionic Hamiltonian
\begin{align}
H &= -t_\mathrm{hop} \sum\limits_{i, j \in \lbrace \mathrm{NN}(i) \rbrace, \sigma } c_{i \sigma}^\dagger c_{j \sigma} + \mathrm{h.c.}  \nonumber \\ &+ J/4 \sum\limits_{i, \alpha \sigma, \sigma^{(1)}, \sigma^{(2)}, \sigma^{(3)}}  c_{i, \sigma}^\dagger  c_{i,\sigma^{(1)}}  f_{i, \sigma^{(2)}}^\dagger f_{i,\sigma^{(3)}}  \tau^\alpha_{\sigma, \sigma^{(1)}}  \tau^\alpha_{\sigma^{(2)}, \sigma^{(3)}},
\end{align}
where $\tau^\alpha_{\sigma, \sigma'}$ are the Pauli matrix components.
We use a Hartree-Fock decomposition of the second term and introduce A/B sublattice magnetizations as well as hybridizations, leading to a self-consistent bilinear Hamiltonian,
\begin{align}
2 m_{f, A/B} &= \left\langle f_{i,A/B, \uparrow}^\dagger f_{i,A/B,\uparrow} \right\rangle - \left\langle f_{i,A/B, \downarrow}^\dagger f_{i,A/B,\downarrow} \right\rangle \\
2 m_{c, A/B} &= \left\langle c_{i,A/B, \uparrow}^\dagger c_{i,A/B,\uparrow} \right\rangle - \left\langle c_{i,A/B, \downarrow}^\dagger c_{i,A/B,\downarrow} \right\rangle \\
V_{A/B} &= \left\langle c_{i, A/B, \uparrow}^\dagger f_{i, A/B \uparrow} \right\rangle  .
\end{align}
For convenience we introduce the total and staggered hybridizations
\begin{align}
V = \frac{1}{2} \left( V_A + V_B \right) \\
V' = \frac{1}{2} \left( V_A - V_B \right).
\end{align}
The self-consistent Hamiltonian reads,
\begin{widetext}
\begin{equation*}
\begin{aligned}
H &= \sum\limits_{\mathbf{k}, \sigma} \underbrace{ \left( c_{\mathbf{k} \sigma A}^\dagger  c_{\mathbf{k} \sigma B}^\dagger f_{\mathbf{k} \sigma A}^\dagger  f_{\mathbf{k} \sigma B}^\dagger  \right) }_{\mathbf{C}^\dagger_{\mathbf{k} \sigma}}  \underbrace{ \begin{pmatrix}
\mu_c + J/2 \sigma m_f & \epsilon( \mathbf{k}  ) & -J/4 (3V - \sigma V') & 0\\
\epsilon( \mathbf{k} ) & \mu_c - J/2 \sigma m_f  & 0 & -J/4 (3V + \sigma V') \\
 -J/4 (3V - \sigma V') & 0 & \mu_f + J/2 \sigma m_c  & 0 \\
 0 & -J/4 (3V + \sigma V') & 0 & \mu_f - J/2 \sigma m_c
\end{pmatrix} }_{\mathbf{H}_{\mathbf{k} \sigma}}
 \underbrace{ 
\begin{pmatrix}
c_{\mathbf{k} \sigma A} \\  c_{\mathbf{k} \sigma B} \\ f_{\mathbf{k} \sigma A} \\  f_{\mathbf{k} \sigma B}
\end{pmatrix}}_{\mathbf{C}_{\mathbf{k} \sigma}} .
\end{aligned}
\end{equation*}
\end{widetext}

\paragraph{Fixing particle number during time evolution}

During the excitation with the laser pulse, the number of $f$ electrons is not necessarily fixed to $1$ per lattice site, due to hybridization with the conduction band electrons. While this issue can be fixed using the chemical potential of the $f$ electrons in equilibrium, in non-equilibrium the problem is more subtle, as the effect of the chemical potential on the $f$ filling shows up only in second order. In the following we derive a method in order to fix the first derivative of the $f$ electron number and thereby also the electron number itself.
The time evolution of an arbitrary bilinear observable $O$ with matrix representation $\mathbf{O}$ can be written as
\begin{align}
O(t) &=  \frac{1}{N} \sum\limits_\mathbf{k} \left\langle \mathbf{D}^\dagger_{\mathbf{k} \sigma} \mathbf{U}^\dagger_{\mathbf{k} \sigma}  \mathbf{V}^\dagger_{\mathbf{k} \sigma}(t)  \mathbf{O} \mathbf{V}_{\mathbf{k} \sigma}(t) \mathbf{U}_{\mathbf{k} \sigma} \mathbf{D}_{\mathbf{k} \sigma} \right\rangle
\end{align}
with
\begin{align}
i \partial_t \mathbf{V}_{\mathbf{k} \sigma}(t) = \mathbf{H}_{\mathbf{k} \sigma} \mathbf{V}_{\mathbf{k} \sigma}(t),
\end{align}
being the Heisenberg time evolution and $ \mathbf{U}_{\mathbf{k}} \mathbf{D}_{\mathbf{k} \sigma} =  \mathbf{C}_{\mathbf{k} \sigma}$ being the transformation to the diagonal operator basis.
The derivative of an Observable that does not explicitly depend on time reads
\begin{align}
i \partial_t O(t) &=  \frac{1}{N} \sum\limits_\mathbf{k} \left\langle \mathbf{D}^\dagger_{\mathbf{k} \sigma} \mathbf{U}^\dagger_{\mathbf{k} \sigma}  \mathbf{V}^\dagger_{\mathbf{k} \sigma}(t)  \left[ \mathbf{O}, \mathbf{H}_{\mathbf{k} \sigma} \right] \mathbf{V}_{\mathbf{k} \sigma}(t) \mathbf{U}_{\mathbf{k} \sigma} \mathbf{D}_{\mathbf{k} \sigma} \right\rangle
\end{align}
We want to fix the f-electron filling factor during the time evolution by adjusting $\mu_f$ in time
\begin{align}
i \partial_t  n_f = \frac{1}{N} \sum\limits_\mathbf{k} \left\langle \mathbf{D}^\dagger_{\mathbf{k} \sigma} \mathbf{U}^\dagger_{\mathbf{k} \sigma}  \mathbf{V}^\dagger_{\mathbf{k} \sigma}(t) \left[ \mathbf{N}_f, \mathbf{H}_{\mathbf{k} \sigma} \right] \mathbf{V}_{\mathbf{k} \sigma}(t) \mathbf{U}_{\mathbf{k} \sigma} \mathbf{D}_{\mathbf{k} \sigma} \right\rangle    
\end{align}
where
\begin{align}
 \mathbf{N}_f = \begin{pmatrix}
 0, 0, 0, 0\\
 0, 0, 0, 0\\
 0, 0, 1, 0\\
 0, 0, 0, 1
 \end{pmatrix}
\end{align}
is the matrix representation of $n_f$.
This first derivative does not depend on $\mu_f$ explicitly, hence we have to go one step further and calculate the second derivative. We first define
\begin{widetext}
\begin{align}
\mathbf{K}_{NH}(t) &=  \left[ \mathbf{N}_f, \mathbf{H}_{\mathbf{k} \sigma} \right] \\
&=  \begin{pmatrix}
0 & 0 & J/4 (3V - \sigma V') & 0\\
0 & 0 & 0 & J/4 (3V + \sigma V') \\
 -J/4 (3V - \sigma V') & 0 & 0  & 0 \\
 0 & -J/4 (3V + \sigma V') & 0 & 0
\end{pmatrix} 
\end{align}
We later need its derivative
\begin{align}
i \partial_t \mathbf{K}_{NH}(t) = i \begin{pmatrix}
0 & 0 & J/4 (3\partial_t V - \sigma  \partial_tV') & 0\\
0 & 0 & 0 & J/4 (3\partial_tV + \sigma \partial_tV') \\
 -J/4 (3\partial_tV - \sigma \partial_tV') & 0 & 0  & 0 \\
 0 & -J/4 (3\partial_tV + \sigma \partial_tV') & 0 & 0
\end{pmatrix} 
\end{align}
it depends on the derivatives of the hybridization on the A/B sublattice, which read
\begin{align}
i \partial_t V_{A/B} &= i \partial_t \frac{1}{2N} \sum\limits_i \left\langle c_{i, A/B, \uparrow}^\dagger f_{i, A/B \uparrow} \right\rangle + \left\langle f_{i, A/B, \uparrow}^\dagger c_{i, A/B \uparrow} \right\rangle \\
 &=\frac{1}{N} \sum\limits_\mathbf{k} \left\langle \mathbf{D}^\dagger_{\mathbf{k} \sigma} \mathbf{U}^\dagger_{\mathbf{k} \sigma}  \mathbf{V}^\dagger_{\mathbf{k} \sigma}(t)  \left[ \mathbf{V}_{A/B}, \mathbf{H}_{\mathbf{k} \sigma} \right] \mathbf{V}_{\mathbf{k} \sigma}(t) \mathbf{U}_{\mathbf{k} \sigma} \mathbf{D}_{\mathbf{k} \sigma} \right\rangle
\end{align}
Now we can calculate the second derivative of the f electron filling factor
\begin{align}
- \partial_t^2  n_f &= \frac{1}{N} \sum\limits_\mathbf{k} & \left\langle \mathbf{D}^\dagger_{\mathbf{k} \sigma} \mathbf{U}^\dagger_{\mathbf{k} \sigma}  \mathbf{V}^\dagger_{\mathbf{k} \sigma}(t) \left[ \mathbf{K}_{NH}(t), \mathbf{H}_{\mathbf{k} \sigma} \right] \mathbf{V}_{\mathbf{k} \sigma}(t) \mathbf{U}_{\mathbf{k} \sigma} \mathbf{D}_{\mathbf{k} \sigma} \right\rangle   \\
&  & +  \left\langle \mathbf{D}^\dagger_{\mathbf{k} \sigma} \mathbf{U}^\dagger_{\mathbf{k} \sigma}  \mathbf{V}^\dagger_{\mathbf{k} \sigma}(t) \left( i \partial_t \mathbf{K}_{NH}(t) \right) \mathbf{V}_{\mathbf{k} \sigma}(t) \mathbf{U}_{\mathbf{k} \sigma} \mathbf{D}_{\mathbf{k} \sigma} \right\rangle 
\end{align}
\end{widetext}
We now determine $mu_f$ in a self-consistent loop at each time step, such that the second derivative of $n_f$ vanishes. As long as the external perturbation does not change $n_f$ in a non-continuous way, this also fixes $n_f=1$.

\paragraph{ Time dependent variational Monte Carlo}  
The variational wave function we employ in VMC and tVMC is given by
\begin{align}
\left| \Psi \right\rangle = P_G P_J \left| \phi \right\rangle, \quad  \left| \phi \right\rangle = \left( \sum\limits_{\alpha, \beta =  \langle c, f \rangle } \sum\limits_{i,j}^{N_\mathrm{s}} f_{ij} \alpha_{i \uparrow}^\dagger \beta_{j \downarrow}^\dagger \right)^{N_e/2} \left| 0 \right\rangle .
\end{align}
Here $\left| \phi \right\rangle$ is the Pfaffian wave function for conduction band and $f$ electrons. The variational parameters in the Pfaffian wave function are given by the $f_{ij}$. The correlation factors $P_G$ and $P_J$ are of Gutzwiller and Jastrow type. For $f$ electrons the Gutzwiller factor is the only relevant correlation factor, as it permits only single occupancy of the lattice sites, effectively casting the $f$ degrees of freedom to a single spin-$1/2$, as relevant for the Kondo lattice model. The variational parameters are subject to a $2 \times 2$ sublattice structure \cite{MISAWA2019447}. We employ the time-dependent variational principle onto the wave function to compute the time dependence of the variational parameters \cite{PhysRevB.88.075133}. For further details to the numerical approach we refer to the supplemental Material and to Ref.\ \cite{PhysRevB.102.165128}.

\section{Acknowledgements} 
We thank Rohit Prasankumar, Filip Ronning and Qimiao Si for helpful discussions. This work was supported by the U.S. DOE NNSA under Contract No. 89233218CNA000001 via the LANL LDRD Program. It was supported in part by the Center for Integrated Nanotechnologies, a U.S. DOE Office of Basic Energy Sciences user facility, in partnership with the LANL Institutional Computing Program for computational resources.

\normalem

\end{document}